%% file: icpm21_main.tex
\documentclass[conference]{IEEEtran}
\IEEEoverridecommandlockouts
\usepackage{cite}
\usepackage{amsmath,amssymb,amsfonts}
\usepackage{algorithmic}
\usepackage{graphicx}
\usepackage{textcomp}
\usepackage{xcolor}
\usepackage{orcidlink}
\usepackage{graphicx}
\usepackage{hyperref}
\usepackage{multicol}
\usepackage[final]{changes}
\usepackage{url}
\usepackage{amsmath}
\usepackage{bm}
\usepackage{multirow}
\usepackage{tcolorbox}
\usepackage{rotating}
\usepackage{latexsym,amssymb,amsmath}
\usepackage{makecell}
\usepackage{xspace}
\usepackage{paralist}
\usepackage{wrapfig}
\usepackage{adjustbox}
\usepackage{cite}
\usepackage{todonotes}
\input{neue_imports}
\input{neue_macros}
\input{macros-style}
\input{macros-common}
\usepackage{etoolbox}
\usepackage{hyperref}
\def\BibTeX{{\rm B\kern-.05em{\sc i\kern-.025em b}\kern-.08em
    T\kern-.1667em\lower.7ex\hbox{E}\kern-.125emX}}
\begin{document}

\title{Probabilistic Trace Alignment\\
\thanks{This research has been partially supported by the project IDEE (FESR1133) funded by the Eur.\ Reg.\ Development Fund (ERDF) Investment for Growth and Jobs Programme 2014-2020. }
}

\author{\IEEEauthorblockN{1\textsuperscript{st} Giacomo Bergami\orcidlink{0000-0002-1844-0851}}
\IEEEauthorblockA{\textit{Dept. of Computer Science} \\
\textit{Free University of Bozen-Bolzano}\\
Bozen-Bolzano, Italy \\
gibergami@unibz.it}
\and
\IEEEauthorblockN{2\textsuperscript{nd} Fabrizio Maria Maggi\orcidlink{0000-0002-9089-6896}}
\IEEEauthorblockA{\textit{Dept. of Computer Science} \\
\textit{Free University of Bozen-Bolzano}\\
Bozen-Bolzano, Italy \\
maggi@inf.unibz.it}
\and
\IEEEauthorblockN{3\textsuperscript{rd} Marco Montali\orcidlink{0000-0002-8021-3430}}
\IEEEauthorblockA{\textit{Dept. of Computer Science} \\
\textit{Free University of Bozen-Bolzano}\\
Bozen-Bolzano, Italy \\
montali@inf.unibz.it}
\and
\IEEEauthorblockN{4\textsuperscript{th} Rafael Peñaloza\orcidlink{0000-0002-2693-5790}}
\IEEEauthorblockA{\textit{--} \\
\textit{University of Milano-Bicocca}\\
Milan, Italy \\
rafael.penaloza@unimib.it}
}

\author{
	\IEEEauthorblockN{
		Giacomo Bergami\orcidlink{0000-0002-1844-0851}\IEEEauthorrefmark{1},
		Fabrizio Maria Maggi\orcidlink{0000-0002-9089-6896}\IEEEauthorrefmark{1},
		Marco Montali\orcidlink{0000-0002-8021-3430}\IEEEauthorrefmark{1},
		Rafael Peñaloza\orcidlink{0000-0002-2693-5790}\IEEEauthorrefmark{2}}

	\IEEEauthorblockA{\IEEEauthorrefmark{1}Free University of Bozen-Bolzano, Bozen, Italy\\
		Email: gibergami@unibz.it, \{maggi,montali\}@inf.unibz.it}
	\IEEEauthorblockA{\IEEEauthorrefmark{2}University of Milano-Bicocca, Milan, Italy\\
		Email: rafael.penaloza@unimib.it}
}

\maketitle

\begin{abstract}
Alignments provide sophisticated diagnostics that pinpoint deviations in a trace with respect to a process model and their severity. However, approaches based on trace alignments use crisp process models as reference and recent probabilistic conformance checking approaches check the degree of conformance of an event log with respect to a stochastic process model instead of finding trace alignments. In this paper, for the first time, we provide a conformance checking approach based on trace alignments using stochastic Workflow nets. Conceptually, this requires to handle the two possibly contrasting forces of the cost of the alignment on the one hand and the likelihood of the model trace with respect to which the alignment is computed on the other.
\end{abstract}


\input{sections/01_introduction}
\input{sections/02_related_works}
\input{sections/03_preliminaries}

\input{sections/06_embedding_proposal}

\input{sections/07_experiments}
\input{sections/08_conclusions}

\bibliographystyle{IEEEtran}
\bibliography{biblio3}

\end{document}

%% file: neue_imports.tex
\usepackage{graphicx}
\usepackage{xcolor,color}
\usepackage{subfig}
\usepackage{tikz}
\usepackage{calc}

\usepackage{tabularx}
\usepackage{booktabs}

\usepackage{ulem}

\usepackage{amsmath,amsfonts}
\usepackage{braket}
\usepackage{xfrac}

\usepackage{pifont}
\usepackage{amssymb}
\usepackage{paralist}
\usepackage[inline]{enumitem}

%% file: neue_macros.tex
\newcommand{\pmin}{\rho}

\newcommand{\const}[1]{\mathsf{#1}}

\newcommand{\alphabet}{\Sigma}
\newcommand{\tasks}{\mathcal{A}}

\newcommand{\uswn}{SWN\xspace}
\newcommand{\net}{\ensuremath{N}}
\newcommand{\tg}{\ensuremath{G}}
\newcommand{\closed}[1]{\overline{#1}}

\newcommand{\prob}[2]{\mathbb{P}_{#2}(#1)}
\newcommand{\rg}[1]{RG(#1)}
\newcommand{\ind}[1]{\textnormal{\texttt{#1}}}
\newcommand{\seq}{\eta}

\newcommand{\trace}{\sigma}

\newcommand{\ptraces}[2]{\mathit{ptraces}_{#2}(#1)}

\newcommand{\seqs}[2]{seqs_{#2}(#1)}

\newcommand{\embed}{\phi}
\newcommand{\trembed}{{\embed^{\text{tr}}}}
\newcommand{\gorgembed}{{\embed^{g}}}

\newcommand{\pa}{\rho_{23}}
\newcommand{\pb}{\rho_{24}}
\newcommand{\pc}{\rho_{55}}
\newcommand{\pd}{\rho_{65}}
\newcommand{\pe}{\rho_{67}}
\newcommand{\pf}{\rho_{57}}

\newcommand{\logtrace}{\trace}
\newcommand{\nonlogtrace}{{\trace'}}

\newcommand{\approptoinn}[2]{\mathrel{\vcenter{
			\offinterlineskip\halign{\hfil$##$\cr
				#1\propto\cr\noalign{\kern2pt}#1\sim\cr\noalign{\kern-2pt}}}}}

\newcommand{\unravelled}{unfolded}

\newcommand{\Ind}[1]{\ind{i}_{#1}}

\newcommand{\TBf}[2]{{\mathbf{G}_{#1}}}
\newcommand{\expN}{\closed{\tg_{\rg{\net}}}}

\def\EqualityHolds{itholds}
\ifdefined\EqualityHolds
\newcommand{\probarg}{\tg}
\newcommand{\WCal}[2]{\ptraces{\tg}{#1}}
\else
\newcommand{\probarg}{\expN}
\newcommand{\WCal}[2]{\ptraces{\closed{\tg_{\rg{\net}}}}{#1}}
\fi
\newcommand{\probskip}[1]{\prob{#1}{\probarg}}
\newcommand{\goldenrank}{\mathcal{R}} 

%% file: macros-style.tex
\newlist{myalist}{enumerate*}{1}
\setlist[myalist]{label=\textbf{(\arabic*)}}
\newlist{mylist}{enumerate*}{1}
\setlist[mylist]{label=\textit{(\roman*)}}
\newlist{alphalist}{enumerate*}{1}
\setlist[alphalist]{label=\textbf{(\alph*)}}

%% file: macros-common.tex
\newcommand{\norm}[2]{\|#1\|_{#2}}

\newcolumntype{C}{>{\centering\arraybackslash}X}

\newcounter{dummy} 
\newcounter{dummy1} 
\newcounter{dummy2}
\newcounter{dummy3} 
\newcounter{dummy4}
\newcounter{dummy5} 
\newcounter{dummy6}

\usepackage[thmmarks,amsmath]{ntheorem}
\theoremseparator{.}
\theorembodyfont{\itshape}
\theoremsymbol{$\triangleleft$}
\newtheorem{theorem}[dummy]{Theorem}
\newtheorem{lemma}[dummy1]{Lemma}
\newtheorem{definition}[dummy2]{Definition}
\newtheorem{proposition}[dummy3]{Proposition}

\theorembodyfont{\normalfont}
\newtheorem{example}[dummy4]{Example}
\newtheorem{remark}[dummy5]{Remark}

\theoremstyle{nonumberplain}
\theoremheaderfont{\itshape}
\theorembodyfont{\normalfont}

\theoremseparator{.}
\theoremsymbol{\ensuremath{\dashv}}
\newtheorem{proof}[dummy6]{Proof}

\qedsymbol{\ensuremath{\dashv}}

%% file: sections/01_introduction.tex
\section{Introduction}

\label{introduction}
%Trace alignment is a well-known technique in conformance checking \cite{DBLP:conf/edoc/AdriansyahDA11} providing both a numerical assessment of the degree of conformance of a log trace with respect to a model, as well as a repair strategy if such trace does not conform to the given model. At the time of the writing,
%
In the existing literature on conformance checking, a common approach is based on trace alignment \cite{DBLP:conf/edoc/AdriansyahDA11}. This approach uses crisp process models as reference models. Yet, recently developed probabilistic conformance checking approaches provide a numerical quantification of the degree of conformance
%the existing approaches are used to check the degree of conformance
of an event log with a stochastic process model by either assessing the distribution discrepancies \cite{DBLP:conf/bpm/LeemansSA19}, or by exploiting entropy-based measures \cite{DBLP:conf/icpm/PolyvyanyyK19,DBLP:journals/tosem/PolyvyanyySWCM20}.
As these strategies are not based on trace alignments, these cannot be directly used to repair a given trace with one of the traces generated by a stochastic process model.
%As traces generated by such models are associated to a probability exhibiting its representativeness and relevance within the model, probabilistic trace alignment techniques should take into account the combined provision of trace probability and alignment cost.
%instead of finding trace alignments.
%
In this paper, we provide for the first time an approach for the probabilistic alignment of a trace and a stochastic reference
model. This approach is not comparable with the existing literature on probabilistic conformance checking as its output is not numeric but consists of a ranked list of alignments.
Providing different alignment options is useful since, conceptually, probabilistic trace alignment requires the analyst to 
%handle the two possibly contrasting forces of the cost of the alignment on the one hand and the likelihood of the model trace with respect to which the alignment is computed.
%We consider the important tradeoff between both
%aspects.
balance between the likelihood of the model trace with respect to which the alignment is computed and the cost of the alignment. 
%(if the cost of the alignment is too high even if the model trace is very likely applying too many changes in the original trace is in turn not very likely).

\begin{figure}[!t]
	\centering
	\includegraphics[width=.49\textwidth]{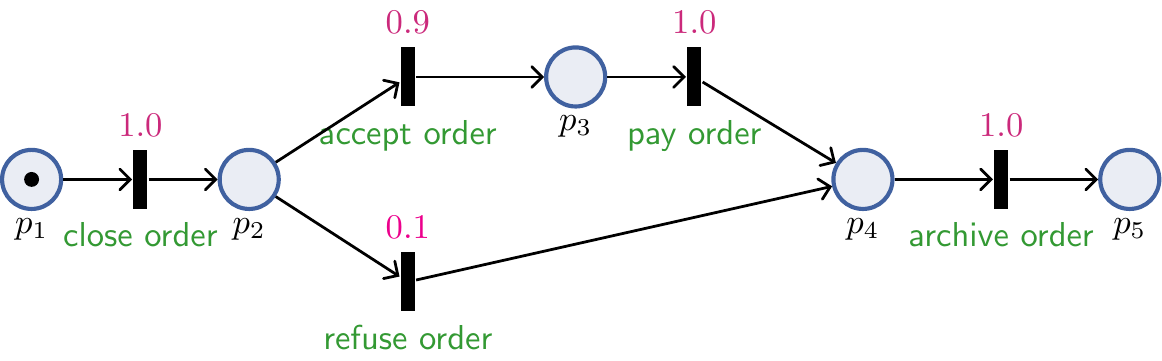}
	\caption{a simple Stochastic Workflow Net.}\label{fig:petri_tut}
	%\vspace{-1.4cm}
\end{figure}
%For example, the probabilistic alignment of a trace with a stochastic net could be represented by the model trace maximizing the combined provision of minimum trace alignment cost and maximum model trace probability. 
With reference to Figure~\ref{fig:petri_tut}, a user might be interested to align the log trace $\langle \textsf{close order},\,\textsf{archive order}\rangle$ with one of the two possible model traces $\langle\textsf{close order},$ $\textsf{accept order},\,\textsf{pay order},\,\textsf{archive order}\rangle$ or $\langle$\textsf{close order}, \textsf{refuse order}, \textsf{archive order}$\rangle$. While the latter trace provides the least alignment cost though the model trace has a low probability ($0.1$), the former gives a slightly greater alignment cost while providing a higher model trace probability ($0.9$). Since, depending on the context, analysts might prefer either the former or the latter alignment, providing a selection of the best $k$ alignments among all the distinct model traces empowers the analysts to find their own trade-off between alignment cost and model trace probability.
%However, in some cases, the user could prefer to identify an alignment with a lower cost even if based on a less probable model trace, while, in other cases, the user could favor a model trace with a higher probability at the expense of a higher alignment cost. Therefore, to provide users with an instrument that allows them to find their own trade-off between alignment cost and model trace probability, we need to return the best $k$ alignments among all the distinct model traces.

%Since when aligning an event log with a stochastic net distinct model traces have different probabilities, the retrieval of the best model trace maximizing the combined provision of minimum trace alignment cost and maximum model trace probability might not suffice. In some cases, indeed, the user could prefer to identify an alignment with a lower cost even if based on a less probable model trace, while, in other cases, the user could favor a model trace with a higher probability at the expense of a higher alignment cost. We consider, therefore, the important tradeoff between both aspects.
%Therefore, in this paper, we propose trace alignment approaches that return the best

To do this, we frame the probabilistic trace alignment problem into the well-known $k$-Nearest Neighbors ($k$NN) problem \cite{Altman} that refers to finding the $k$ nearest data points to a \textit{query}  from a set  of \textit{data points} via a distance function.
We introduce two ranking strategies. The first one is based on a brute force approach that reuses existing trace aligners such as \cite{DBLP:conf/edoc/AdriansyahDA11,LeoniM17}, where the (optimal) ranking of the top-k alignments is obtained by computing the Levensthein distance between the trace to be aligned and all the model traces and by multiplying each of these distances by the probability of the corresponding model trace. However, even if this approach returns the best trace alignment ranking for a query trace, the alignments must be computed a-new for all the possible traces to be aligned. For models generating a large number of model traces, this would clearly become unfeasible. Therefore, we propose a second strategy that produces an approximate ranking where traces are represented as numerical vectors via an embedding. {Then, by exploiting ad-hoc data structures,
	%such as Vp-Trees \cite{Fu2000}, Kd-Trees \cite{Maneewongvatana99}, and M-Trees \cite{Ciaccia},
	we can retrieve the neighborhood of size $k$ containing the traces similar to the given query  by pre-ordering (\textit{indexing}) the model traces  via the aformentioned distance. 
	%Thus, we do not need to analyze the entire space, but just start the search from the top-$1$ alignment. 
	If the embeddings for our model traces are independent of the query of choice, this would not require to constantly recompute the numeric vector representation for the model traces.
	%	
	
	%%%%% Proposed part as the last part of the introduction:
	%\texttt{\color{red}[TODO]}
	%\todo{this is too specific for an introduction; in particular, too many details on how the experiments are done.}
	We implemented both strategies and perform experiments using a real life event log coming from a hospital system to empirically evaluate the properties of our proposed  strategy. We assessed our proposed as follows:
\begin{mylist}
	\item first, we evaluate the degree of approximation introduced by the approximate-ranking approach if compared with the optimal-ranking. We observe that different embedding strategies provide a trade-off between ranking stability vs. precision (\S\ref{subsec:apprp}).
	\item Last, we evaluate the computational time required to both generate the embeddings and to assess the similarity between the embeddings. We observe that approximate-ranking alignments provide the best trade-off between accuracy and efficiency (\S\ref{subsec:efficio}).
\end{mylist}

\begin{figure*}[!t]
	\begin{minipage}{.49\textwidth}
		\centering
		\includegraphics[width=.7\textwidth]{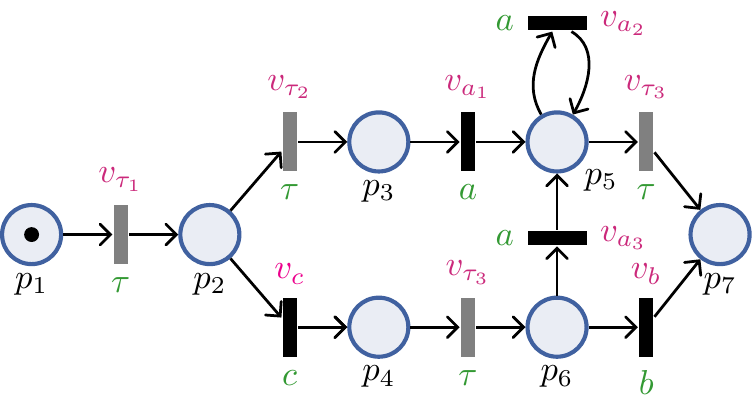}
		\caption{A sample \uswn $N$. Labels are shown in green, $\tau$ transitions in grey, weights in magenta.}\label{fig:spn}
	\end{minipage}\hfill \begin{minipage}{.49\textwidth}
	\centering
		\includegraphics[width=.7\textwidth]{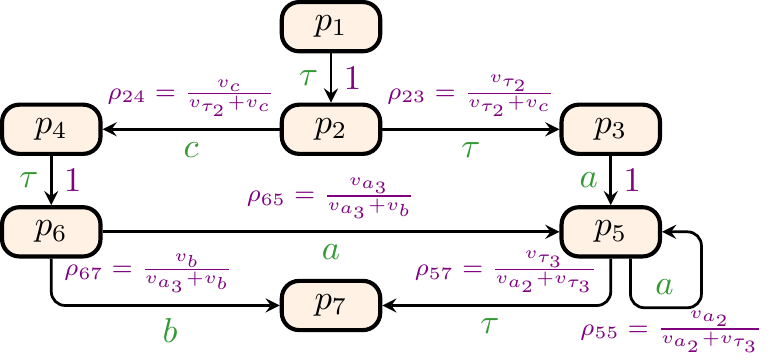}
		\caption{Reachability graph  of the \uswn $N$. Probabilities are shown in violet.}\label{fig:rg}
	\end{minipage}

	\begin{minipage}{.4\textwidth}
		\centering \includegraphics[width=.7\textwidth]{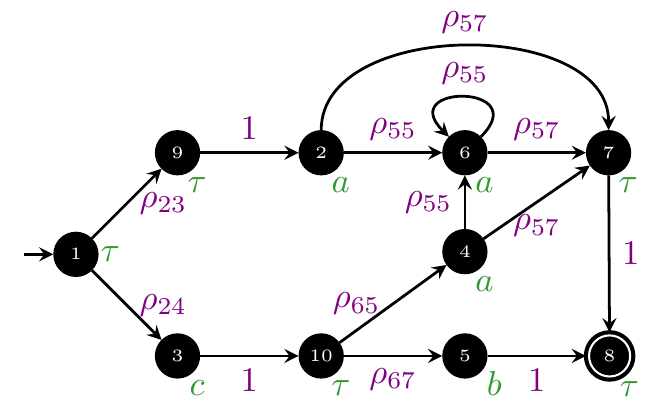}
		\caption{Preliminary Transition graph $TG$ encoding the SWN $N$ with no $\tau$-closures.}\label{fig:lmc}\label{fig:orig}
	\end{minipage}\hfill \begin{minipage}{.4\textwidth}
	\centering \includegraphics[width=.7\textwidth]{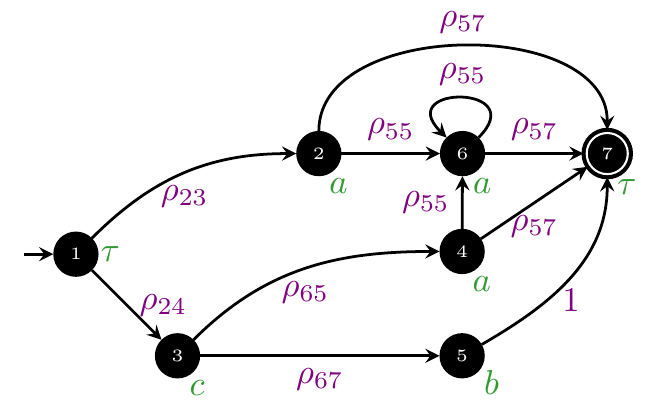}
		\caption{Transition graph $TG$ resulting from $N$ after $\tau$-closure.}\label{fig:closed}
	\end{minipage}
	\vspace{-.6cm}
\end{figure*}

%% file: sections/02_related_works.tex
\section{Related Works}
%%%
%%%
\paragraph*{Stochastic Conformance Checking} earlier works on probabilistic trace alignments \cite{AlizadehLZ14a} extended alignment cost functions by considering probabilities of never occurring activities when activities on both traces do not match. Such functions always return a zero cost alignment cost on identical matching traces, independently from the probability associated to the model trace. This approach favors traces providing optimal alignment cost: the resulting ranking cannot be used to provide a trade-off between trace probability and alignment cost. As the model has non-stochastic Petri Net as a reference model, the computation of the aforementioned probability for activities' mismatch requires a log file for providing such an estimation as the more recent work \cite{spdwe}, while our proposed solution might as well estimate trace probability by directly loading Stochastic Petri Nets. We  deduce that our proposed solution proves to be more general that this first attempt to probabilistic trace alignments. 

More recent works on stochastic model checking  assess the degree of conformance of the whole stochastic model against either one single log trace \cite{DBLP:conf/icpm/PolyvyanyyK19,DBLP:journals/tosem/PolyvyanyySWCM20} or an entire log \cite{LeemansSA19}, and consider Stochastic Petri Nets. %On the other hand, we are interested in determining which is the best model trace providing the trade-off between trace probability and alignment cost with the log trace to be aligned, thus limiting the model probability distribution to the part generating one sole given trace at a time. Therefore, such 
With respect to the formers, we might rank models according to the its degree of conformance of a fixed log trace while, on the other hand, our proposed solutions ranks a subset of the possible model traces according to a given trace. Albeit the input of such approaches is the same as ours, the problem that we intend to solve is different. Nevertheless, we might exploit our solution to rank stochastic models via the best trace alignment provided by each single model.  Furthermore, approaches considering entire logs as a whole cannot be possibly exploited for aligning one stochastic model to a single log trace \cite{LeemansSA19}, as for our assumptions the remaining log traces remain unknown, and therefore it is impossible to ``earth-move'' probability distribution from a stochastic model towards a set of unknown traces except one. Our approximated ranking approach reduces the alignment problem to the computation of the Euclidean Distance independently from the log trace that we want to align, thus avoiding the cost of repeatedly extracting the model traces and providing an ad-hoc vectorial representation. Such re-computation is still doable in our optimal ranking representation.

%%%
%%%
\paragraph*{Graph Kernels} graph kernels express similarity measures \cite{Samatova} involved in both classification \cite{TsudaS10} and clustering algorithms. One of the first approaches required a preliminary embedding definition of topological description vectors extracted from the most frequent subgraphs within a graph database \cite{Sidere}. As a drawback, it required the computation of a subgraph isomorphism problem, which is NP-complete. In fact, the definition of a graph kernel function fully recognizing the structure the graph always boils down to solving such NP-Complete problem \cite{GartnerFW03}, as exact embeddings generable in polynomial can be inferred just for loop-free Direct Acyclic Graphs \cite{BergamiBM20}. Consequently, most recent literature focused on extracting relevant features of such graphs, that are then used to define a graph similarity function. The most common approach adopted in the kernel to extract such features is called \textit{propositionalization}: we might extract all the possible features (e.g., subsequences), and then define a kernel function based on the occurrence and similarity of these features \cite{Gartner03}. For node-labelled graphs, the features come from the node labels and the possible strings that might be generated while traversing the graph (see \cite{Gartner03} and \S\ref{subsec:katk}). 

%% file: sections/03_preliminaries.tex
\section{Foundational Components and Assumptions}
We start by introducing the foundational components of our approach, and the corresponding working assumptions. On the one hand, we describe the class of  Petri nets we consider to represent stochastic processes. On the other hand, we recall graph and string kernels, which we use to compute probabilistic alignments.

\subsection{Stochastic Workflow Nets}\label{subsec:spn}
To model stochastic processes, we isolate an interesting class of Stochastic Petri nets \cite{MarsanCB84,Desel1998,RoggeSoltiAW13}. From the structural point of view, we consider $k$-bounded workflow nets with silent transitions. We assume a set $\alphabet = \tasks \cup \set{\tau}$ of labels, where labels in $\tasks$ indicate process tasks, whereas $\tau$ stands for an invisible execution step ($\tau$-transition). %Labels are associated to transitions via a labelling function $\lambda$.
From the stochastic point of view, do not consider timed aspects, and only focus on the definition of a probability distribution over enabled transitions. We call these nets \emph{stochastic workflow nets} (SWN for short)
Formally, an SWN is a tuple $\net=(P,T,F,W,i,f)$ where:
\begin{mylist}
	\item $P$ is a set of \textit{places};
	% to which we can associate a finite number of indistinguishable tokens;
	\item $T$ is a set of \textit{transitions} $t\in T$, to which we associate a label $\lambda(t)\in\Sigma$;
	\item $F\subseteq (P\times T)\cup (T\times P)$ is the \emph{flow relation} linking places to transitions and transitions to places; % to which we associate a \textit{firing cost} $\omega\colon F\to\mathbb{N}$;
	\item $W\colon T\to \mathbb{R}$ defines a \textit{firing weight} associated to each transition;
	\item the \emph{initial place} $i\in P$ has no ingoing edges; %($\not\exists t\in T. (t,i)\in F$);
	\item the \emph{final place} $f\in P$ has no outgoing edges. %($\not\exists t\in T. (f,t)\in F$).
\end{mylist}
The notions of marking, transition enablement, transition firing, and reachable markings, are as usual.
%A \textit{marking} is an assignment of a given amount of indistinguishable tokens to places described by a vector $M\colon P\to \mathbb{N}$. We say that a given transition $t$ is \textit{enabled} if $M(p)\geq \omega(p,t)$ for each ingoing $p$ to $t$ ($(p,t)\in F$). If such transition is enabled, then it can \textit{fire} a token. The set of the \textit{enabling transitions} $E(M)$ for a given marking $M$ are all the $t$ reachable from $p$ ($(p,t)\in F$) with $M(p)\neq 0$ where $t$ is enabled. When $t$ can fire a token for a marking $M$, we can generate a novel marking $M'$ from $M$ by moving the tokens from the ingoing places towards the outgoing places as
%$\forall p\in P.\; M'(p)=M(p)-[\omega(p,t)]+[\omega(t,p)]$.

We denote by $E(M)$ the set of transitions of $\net$ that are enabled in $M$. Given an initial marking $M$ over $\net$,  the \textit{reachability graph} of $(\net,M)$ is a graph $(\mathcal{M},\mathcal{E})$ with nodes $\mathcal{M}$ and $T$-labelled edges $\mathcal{E}$, where nodes $\mathcal{M}$ are all the reachable markings from $M$ ($M$ included), and where there is an edge from $M$ to $M'$ labeled by $t$ iff $t \in E(M)$ and firing $t$ in $M$ produces $M'$ (this is indicated by $M\overset{t}{\to}M'$).

We consider the two special markings $M_i$ and $M_f$ of $\net$, respectively assigning one token to the initial place $i$ and the final place $f$, and no tokens elsewhere. We say that $\net$ is $k$-\emph{bounded} if each marking in the reachability graph of $(\net,M_i)$ does assigns at most $k$ tokens to each place of $\net$. \emph{As a first, working assumption, we concentrate on $1$-bounded (i.e., safe) nets} (but our technique seamlessly carries over $k$-bounded nets as well).

As usual in stochastic nets, given a marking $M$ we use $W$ to induce a probability distribution over $E(M)$. This is done by assigning to each edge $M\overset{t}{\to}M'$ a corresponding transition probability $\mathbb{P}\left(M\overset{t}{\to}M'\right)=\frac{W(t)}{\sum_{t'\in E(M)}W(t')}$ \cite{spdwe}. Notice that, by construction, the probabilities associated to all enabled transitions in a marking always add up to 1.

A \emph{run} $\seq$ over $\net$ is a finite sequence $t_1\cdots t_n$ of transitions leading from $M_i$ to $M_f$ in the reachability graph of $(\net,M_i)$. The probability $\prob{\seq}{\net}$ of $\seq$ is obtained as as the product of the probabilities associated to each transition via $W$.

\begin{remark}
\label{rem:collective}
The sum of the probabilities of all runs of a SWN, up to a certain maximum length $n$, is always between 0 and 1. When $n$ tends to $\infty$, this sum tends to $1$. This is a direct consequence of how transition probabilities are computed, paired with the fact that runs of a workflow net are maximal.
\end{remark}

%To each of such edges $M\overset{t}{\to}M'$, we associate a transition probability. We say that a SPN with initial marking $M$ is $k$-\textit{bounded} if each of the nodes $M$ in the reachability graph have $\forall p\in P.\; M(p)\leq k$.

We now turn to traces, defined as finite sequence of labels from $\tasks$ (where each element witnesses the execution of a visible activity). The main issue is that, due to $\tau$-transitions, traces do not directly map into runs. To distinguish model traces from all the possible traces $\tasks^*$, we then proceed as follows. Following \cite{DBLP:conf/edoc/AdriansyahDA11,LeoniM17},  we say that a trace $\nonlogtrace=\const{a}_1\cdots \const{a}_m$ is a \emph{model trace for $\net$} ($\net$-trace for short) if there exists a run $\seq = t_1\cdots t_n$ where the sequence $\lambda(t_1)\cdots \lambda(t_n)$ of labels of $\seq$ coincides with $\nonlogtrace$ once all the $\tau$-labels are stripped away.

There may be multiple, possibly infinitely many runs yielding the same $\net$-trace. Given a $\net$-trace $\nonlogtrace$, we denote by $\seqs{\nonlogtrace}{\net}$ the set of such runs. The probability $\prob{\nonlogtrace}{\net}$ of $\nonlogtrace$ is then obtained by summing up the probabilities of all runs from $\seqs{\nonlogtrace}{\net}$.
This corresponds to the intuition that, to observe $\nonlogtrace$, one can equivalently pick any of its underlying runs. Notably, if a trace is not a $\net$-trace (i.e., it does not conform with $\net$), then its probability is 0.

To make this probability amenable to computation, we apply a \emph{second working assumption, introduced in \cite{Bergami21}, namely that of bounded silence}. A SWN has \emph{bounded silence} if there exists a fixed bound $b$ limiting the maximum number of consequent $\tau$-transitions between two visible transitions of the run. In \cite{Bergami21}, it is argued that this assumption is reasonable when modeling business processes: it retains the possibility of capturing gateways and skippable tasks, while forbidding whole cycles entirely characterized by $\tau$ transitions, which would in turn generate infinitely many runs yielding the same trace.

\begin{remark}
\label{rem:silence}
	For a SWN $\net$ with silence bounded by $b$, given a $\net$-trace $\trace$ there are boundedly many runs of $\net$ yielding $\trace$, that is, $\seqs{\nonlogtrace}{\net}$ has a size bounded by $b$, containing runs whose maximum length is bounded by the length of $\trace$ and $b$ \cite{Bergami21}.
\end{remark}

By combining Remarks~\ref{rem:collective} and \ref{rem:silence}, we thus get a direct way of computing the trace probability $\prob{\nonlogtrace}{\net}$.

Since our goal is to handle probabilistic trace alignment, we need to relate a (possibly non-model) arbitrary trace $\logtrace$ over $\tasks^*$ with the most closely $\net$-traces that balance their distance from $\logtrace$ and their probability. In this respect, we notice the following.

\begin{remark}
By increasing the length of $\net$-traces, we reach a point where their probability and distance w.r.t.~the log trace $\trace$ of interest \emph{both} decrease. Intuitively, this is because executing too many loop iterations within $\net$ decreases the overall run probability, and increments the distance to $\trace$.
\end{remark}

Thanks to this final remark, we get that the problem of probabilistic trace alignment works in a finite space of traces and runs, and it is consequently a combinatorial problem that can be attacked with techniques such as $k$-nearest neighbors.

\subsection{Graph and String Kernels}\label{subsec:katk}
 As a foundational basis to compute trace alignments, we adapt similarity measures from the database literature.  Given a set of data examples $\mathcal{X}$, (e.g., strings or traces, transition graphs) a (positive definite) \emph{kernel} function $k\colon \mathcal{X}\times \mathcal{X}\to \mathbb{R}$ denotes the similarity of elements in $\mathcal{X}$. If $\mathcal{X}$ is the $d$-dimensional Euclidean Space $\mathbb{R}^d$, the simplest kernel function is the inner product $\Braket{\mathbf{x},\mathbf{x}'}=\sum_{1\leq i\leq d}\mathbf{x}_i\mathbf{x}'_i$.
A kernel is said to \emph{perform ideally} \cite{Gartner03} when $k(x,x')=1$ whenever $x$ and $x'$ are the same object (\textit{strong equality}) and $k(x,x')=0$ whenever $x$ and $x'$ are distinct objects (\textit{strong dissimilarity}). A kernel is also said to be \emph{appropriate} when similar elements $x,x'\in\mathcal{X}$ are also close in the feature space. Notice that appropriateness can be only assessed  empirically \cite{Gartner03}.
A positive definite kernel induces a distance metric as
$
d_k(\mathbf{x},\mathbf{x}'):=\sqrt{k(\mathbf{x},\mathbf{x})-2k(\mathbf{x},\mathbf{x}')+k(\mathbf{x}',\mathbf{x}')}
$.
When the kernel of choice is the inner product, the resulting distance is the Euclidean distance $\norm{\mathbf{x}-\mathbf{x}'}{2}$. A normalized vector $\hat{\mathbf{x}}$ is defined as $\mathbf{x}/\norm{\mathbf{x}}{2}$. For a normalized vector we can easily prove that: $\norm{\hat{\mathbf{x}}-\hat{\mathbf{x}}'}{2}^2=2(1-\Braket{\hat{\mathbf{x}},\hat{\mathbf{x}}'})$.
When $\mathcal{X}$ does not represent directly a $d$-dimensional Euclidean space, we can use an \emph{embedding} $\embed\colon\mathcal{X}\to \mathbb{R}^d$ to define a kernel $k_\embed\colon \mathcal{X}\times \mathcal{X}\to\mathbb{R}$ as $k_\embed(x,x'):=\Braket{\embed(x),\embed(x')}$. As a result, $k_\embed(x,x')=k_\embed(x',x)$ for each $x,x'\in\mathcal{X}$.

 The literature also provides a kernel representation for strings \cite{GartnerFW03}: if we associate each dimension in $\mathbb{R}^d$ to a different sub-string $\alpha\beta$ of size $2$ (i.e., $2$-grams\footnote{\label{fn:caveat}For our experiments, we choose to consider only $2$-grams, but any $p$-grams of arbitrary length $p\geq 2$ might be adopted \cite{Gartner03}. An increased size of $p$ improves precision but also incurs in a worse computational complexity, as it requires to consider all the arbitrary subtraces of length $p$ whose constitutive elements occur at any distance from each other within the trace.}), it should represent how frequently and ``compactly'' this subtrace is embedded in the trace $\trace$ of interest. Therefore, we introduce a \emph{decay factor} $\lambda\in[0,1]\subseteq\mathbb{R}$ that, for all $m$ sub-strings where $\alpha$ and $\beta$ appear in $\trace$ at the same relative distance $l < |\trace|$, weights the resulting embedding as $\lambda^lm$.

\begin{table*}[t!]
	\vspace{+0.5cm}
	\caption{Embedding of traces $\const{caba}$, $\const{caa}$ and $\const{cb}$.}\label{tb:embedding}
	\vspace{-0.4cm}
	\begin{center}
		%\scalebox{0.6}
		{
			\begin{tabularx}{\textwidth}{
					>{\hsize=.1\hsize}X
					>{\hsize=.2\hsize}X
					>{\hsize=.1\hsize}X
					>{\hsize=.1\hsize}X
					>{\hsize=.1\hsize}X
					>{\hsize=.1\hsize}X
					>{\hsize=.1\hsize}X
					>{\hsize=.25\hsize}X
					>{\hsize=.2\hsize}X
					>{\hsize=.1\hsize}X
				}
				\toprule
				& $\const{aa}$    & $\const{ab}$   & $\const{ac}$    & $\const{ba}$   & $\const{bb}$   & $\const{bc}$ & $\const{ca}$ & $\const{cb}$ & $\const{cc}$   \\
				\midrule
				$\const{caba}$ & $\lambda^2$ & $\lambda$ & $0$ & $\lambda$  & $0$  & $0$ & $\lambda+\lambda^3$ & $\lambda^2$ & $0$\\
				%$\const{caaa}$ & $2\lambda+\lambda^2$& $0$ & $0$ & $0$ & $0$ & $0$ & $\lambda+\lambda^2+\lambda^3$ & $0$ & $0$ \\
				$\const{caa}$  & $\lambda$ & $0$ & $0$ & $0$ & $0$ & $0$ & $\lambda+\lambda^2$ & $0$&  $0$\\
				$\const{cb}$   & $0$ & $0$ & $0$ & $0$ & $0$ & $0$ & $0$ & $\lambda$& $0$ \\
				\bottomrule
			\end{tabularx}
		}
		\vspace{-0.3cm}
	\end{center}
\end{table*}
\begin{example}\label{ex:wheredotiszero} %\small
	Consider tasks $\tasks=\Set{a,b,c}$. The possible 2-grams over $\tasks$ are $\tasks^2=\Set{\const{aa},\const{ab},\const{ac},\const{ba},\const{bb},\const{bc},\const{ca},\const{cb},\const{cc}}$. Table~\ref{tb:embedding} shows the embeddings of some traces. Being a 2-gram, trace $\const{cb}$ has only one nonzero component, namely that corresponding to itself, with $\trembed_{\const{cb}}(\const{cb})=\lambda$. Trace $\const{caa}$ has the 2-gram $\const{ca}$ occurring with length $1$ ($\const{\underline{ca}a}$) and $2$ ($\const{\underline{c}a\underline{a}}$), and the 2-gram $\const{aa}$ with occurring length $1$ ($\const{c\underline{aa}}$). Hence: $\trembed_{\const{ca}}(\const{caa})=\lambda+\lambda^2$ and  $\trembed_{\const{aa}}(\const{caa})=\lambda$.  Similar considerations can be carried out for the other traces in the table.
	We now want to compute the similarity between the first trace $\const{caba}$ and the other two traces. To do so, we sum, column by column (that is, 2-gram by 2-gram) the product of the embeddings for each pair of traces. We then get $k_{\trembed}(\const{caba},\const{caa})=\lambda^3+(\lambda+\lambda^3)(\lambda+\lambda^2)$ and $k_{\trembed}(\const{caba},\const{cb})=\lambda^3
	$,
	%{\footnotesize
	%\[
	%k_{\trembed}(\const{caba},\const{caaa})=\lambda(\lambda+\lambda^2+\lambda^3)
	%~~
	%k_{\trembed}(\const{caba},\const{caa})=\lambda(\lambda+\lambda^2)
	%~~
	%k_{\trembed}(\const{caba},\const{cb})=\lambda(\lambda+\lambda^3)
	%\]}
	which induces ranking $
	k_{\trembed}(\const{caba},\const{caa})>
	k_{\trembed}(\const{caba},\const{cb})
	$.
\end{example}

Nevertheless, such string embedding has several shortcomings: \begin{alphalist}
	\item it is not weakly-ideal, so we cannot numerically assess if two embeddings represent equivalent traces
	(Example \ref{ex:wheredotiszero});
	\item it does not characterize $\tau$-moves, so the probabilities of the initial and final $\tau$-moves are not preserved; and
	\item it is affected by numerical errors from finite arithmetic: longer traces $\nonlogtrace$ generated from skewed probability
	distributions %$G.\Lambda^i$
	yield greater truncation errors, as smaller $\lambda^i$ components for bigger
	$i<|\nonlogtrace|$ are ignored, preventing a complete numerical vector characterization of  $\nonlogtrace$ in practice.
\end{alphalist}

%% file: sections/06_embedding_proposal.tex
\section{Alignment Strategy}\label{subsec:as}

Starting from common assumptions from the BPM community, we are able to transform reachability graphs into a node-labeled Markovian Process, where  $R$ is the associated transition matrix and $L$ is the node-labeling matrix. In fact, such matrices can determine the probability of reaching a node labeled $\beta\in\Sigma$ from any node labelled $\alpha\in\Sigma$ in $n$ steps as in string kernels:   $[\Lambda^n]_{\alpha\beta}:=[LR^nL^\top]_{\alpha\beta}/[LL^\top]_{\alpha\alpha}$ (see \cite{GartnerFW03} and Example \ref{ex:wheredotiszero}).

First, we need to shift labels from edges to nodes, and then we perform $\tau$-closures while preserving (if required) $\tau$-transitions for both start and final nodes; such operations preserve the traces' probabilities \cite{Bergami21}. Next, the transition probabilities are going to be stored in a matrix $R$ representing a stochastic process, while for each node $i$ there is only one label $\alpha$ such that $[L]_{i\alpha}=1$, and $[L]_{i\beta}=0$ if $\alpha\neq\beta$. Therefore, runs for Markovian Processes are then defined as for SWNs, and we employ the same notation to indicate the valid sequences underlying a model trace. The computation of probabilities for traces is hence defined equivalently. We denote these pair of matrices as \textsc{Transition Graph} $TG=(L,R)$.

When aligning a log trace with a TG, retrieving the model trace maximizing the combined provision of minimum trace alignment cost
and maximum model trace probability does not suffice.
Hence, we find the best $k$ alignments among all  model traces in $\ptraces{\net}{\pmin}$. This reduces to the
$k$-nearest neighbors ($k$NN) problem by finding the $k$ nearest data points to a \textit{query} $x$ from a set
$\mathcal{X}$ of \textit{data points} w.r.t.\ a given distance function $d_k$. Through ad-hoc data structures, such as VP-Trees
and KD-Trees, %VP-Trees \cite{Fu2000} \cite{Maneewongvatana99}, and M-Trees \cite{Ciaccia},
we can retrieve the $k$-neighborhood of $x$ in $\mathcal{X}$ by pre-ordering (\textit{indexing}) $\mathcal{X}$ with respect to $d_k$.
%and searching from the top-$1$ alignment.
%
%To align a trace $\logtrace$ over the \unravelled\ traces $\ptraces{\tg}{\pmin}$,
%the $k$-Nearest Neighbors describe the best $k$ alignments for $\logtrace$.
%We discuss two strategies to obtain these
%alignments.

\smallskip
\noindent
\textbf{Optimal-Ranking Trace Aligner.}
One approach is to reuse existing trace aligners and compute the alignment cost for each model trace $\nonlogtrace$ to be aligned with a model trace at a time $\logtrace$. Customary alignments \cite{DBLP:conf/edoc/AdriansyahDA11,LeoniM17} can be efficiently computed via string Levenshtein distance $d(\logtrace,\nonlogtrace)$: therefore, we will consider traces as strings with associated probability values. Given that similarity is the inverse function of distance and that, concerning the alignment task, we want to find the model trace maximizing both probability and similarity with the log trace, the problem boils down to maximizing the product $p\cdot s$, where $p=\prob{\logtrace}{\mathcal{L}}\prob{\nonlogtrace}{\net}=\prob{\nonlogtrace}{\net}$ and $s=\frac{1}{\frac{1}{c}d(\logtrace, \nonlogtrace)+1}$, where $c\in\mathbb{N}_{\neq 0}$ is a constant. We refer to $p\cdot s$ as the golden ranking function denoted as $\mathcal{R}(\logtrace,\nonlogtrace)$.

\begin{table}[!t]
	\vspace{5mm}
	\centering
	\caption{Golden ranking of model traces with maximum length $4$, where $\logtrace=\const{caba}$ and $c=5$.}\label{tab:expected}
	\resizebox{.45\textwidth}{!}{\begin{tabular}{lc|ll|c}
			\toprule
			
			{$\nonlogtrace$} &
			{$d(\sigma',\sigma)$} &
			$( \mathbb{P}_N(\sigma')$ &  $,\,s_d(\sigma',\sigma)) $ &
			{$\approx s_d(\sigma,\sigma^*)\cdot w_\sigma$} \\

			\midrule
			$\const{a}$  & $3$ & $0.4$ & $\;\; 0.6250$  & $0.2500$\\
			$\const{aa}$  & $2$ & $0.2$ & $\;\; 0.7142$ & $0.1428$\\
			$\const{aaa}$  & $2$ & $0.1$ & $\;\; 0.7142$ & $0.0714$ \\
			$\const{ca}$  & $2$ & $0.07$ & $\;\; 0.7142$ & $0.0500$\\
			$\const{cb}$  & $2$ & $0.06$ & $\;\; 0.7142$ & $0.0428$ \\
			$\const{aaaa}$  & $3$ & $0.05$ & $\;\; 0.7142$ & $0.0357$ \\
			$\const{caa}$  & $1$ & $0.035$ & $\;\; 0.8333$ & $0.0292$ \\
			$\const{caaa}$  & $1$  & $0.0175$ & $\;\; 0.8333$ & $0.0145$ \\
			\bottomrule
	\end{tabular}}
\end{table}
\begin{example}\label{ex:rankingTaus}
	Consider the TG $\closed{\tg}$ in \figurename~\ref{fig:lmc} with probabilities
	$\pa=0.8$, $\pb=0.2$, $\pc=\pf=0.5$, $\pd=0.7$, and $\pe=0.3$. The traces with maximum length $4$ are:
	%$$\begin{aligned}
	%\ptraces{\expN}{0}_{|\nonlogtrace|\leq 4}=
	$\{\braket{\const{a},0.4},\braket{\const{aa},0.2}$, $\braket{\const{aaa},0.1}$, $\braket{\const{ca},0.07}$,
	$\braket{\const{cb},0.06}$,
	$\braket{\const{aaaa},0.05},\braket{\const{caa},0.035},\braket{\const{caaa},0.0175}\}$.
	Table \ref{tab:expected} represents their alignment raking with  $\nonlogtrace=\textup{caba}$.  Although $\const{caa}$ and
	$\const{caaa}$ are the most similar to $\const{caba}$, their associated probability  is rather low, so traces with
	higher probability but lower similarity score are preferred (e.g., $\const{a}$ and $\const{aa}$).
\end{example}
{Since users might still prefer the most similar traces to the ones maximizing both probability and similarity, we
	return the best $k$ solutions.
	We reduce the problem to $k$NN over the Euclidean Space through a transformation $t$ such that the distance of the
	transformed point $t(p,s)$ from  the origin $\vec{0}$ is $\sfrac{1}{ps}$. This preserves the score from the golden ranking
	and maps the points maximizing $ps$ close to the origin, but it requires to recompute the transformation for each new log trace $\logtrace$. We choose}
$t(p,s):=\left(\frac{1}{s\sqrt{p^2+s^2}},\; \frac{1}{p\sqrt{p^2+s^2}}\right)$.
Our search always starts from the origin, and hence finds the best candidates first.
%
%\begin{example}
	{Given a family of hyperbolae $p\cdot s$ with all the alignments having
		$\mathcal{R}(\sigma,\sigma')=p\cdot s$.
%		Point $(1,1)$ is the best possible trace match, i.e., a trace
%		$\nonlogtrace\in\ptraces{G}{0}$ with $\nonlogtrace=\logtrace$ and $\mathbb{P}_G(\nonlogtrace)\mathbb{P}_G(\logtrace)=1$.
%		%		
%		Figure \ref{fig:knnspace} (below) shows that
%		
	the embedding moves the points of the hyperbola $p\cdot s$ to a circumference $x^2+y^2=\sfrac{1}{(ps)^2}$ describing a locus of the points equidistant as $\sfrac{1}{ps}$ from the origin of the axes $(0,0)$.}
%\end{example}
%
The transformation required for running the $k$NN algorithm preserves the golden ranking.

\begin{lemma}
	\label{lem:transfspace}
	The set of points having the product $ps$ at least $k\in[0,1]$ corresponds to the set of $t$-transformed points with distance
	at least $1/k$ from the origin.
\end{lemma}

\noindent
\textbf{Approximate-Ranking Trace Embedder.}\label{subsec:ate}
Ranking optimality comes at the cost of a brute-force recomputation of $\goldenrank$ for each trace $\logtrace$ to align. Alternatively, we might avoid the brute-force cost by
%Each embedding $\phi$ entails an associated similarity metric $k_\phi$ (\S\ref{subsec:katk}) and an associated
%distance $d_{k_\phi}$ (Equation \ref{eq:dofk}), hence
%we can compute
computing the embeddings for all the \unravelled\ traces
before the top-$k$ search ensuring that they are independent of the trace to align. %, avoiding the brute-force cost.
%As previously observed,
This computational gain comes with a loss in precision, that our embedding proposal tries to mitigate by overcoming some of the current literature shortcomings:
%; the generation of precise embeddings for graph data with loops is
%NP-complete \cite{GartnerFW03}. %and, in its approximated version, is unable to accurately represent data using low-dimensional
%%vectors \cite{Seshadhri5631}.
%Our proposed embedding ($\gorgembed$) is thus weakly-ideal (\S\ref{subsec:katk}).
%%
%$\gorgembed$ is a variant of the embedding $\trembed$ from \cite{LodhiSSCW02}, which addresses some of its shortcomings.
%Indeed, $\trembed$
%\begin{alphalist}
%	\item is not weakly-ideal, so we cannot numerically assess if two embeddings represent equivalent traces
%	(Example \ref{ex:wheredotiszero});
%	\item does not characterize $\tau$-moves, so the probabilities of the initial and final $\tau$-moves are not preserved; and
%	\item is affected by numerical errors from finite arithmetics: longer traces $\nonlogtrace$ generated from skewed probability
%	distributions $G.\Lambda^i$ yield greater truncation errors, as smaller $\lambda^i$ components for bigger
%	$i<|\nonlogtrace|$ are ignored, preventing a complete numerical vector characterization of  $\nonlogtrace$ in practice.
%\end{alphalist}
%%
%To overcome these shortcomings we
	 propose a weakly-ideal embedding
 preserving probabilities from and to $\tau$ transitions, and
 mitigating the numerical truncation errors induced by trace length and probability distribution skewness through two
sub-embedding strategies, $\epsilon$ for transition correlations in $\tasks^2$ and $\nu$ for transition label frequency in $\tasks$\cite{Bergami21}.

In order to meet the goal, we need to first transform a SWN into a TG: first, for each model trace $\nonlogtrace$ we need to restrict the TG into a \textit{weighted} TG $G_\nonlogtrace=(TG_{\nonlogtrace},\omega_\nonlogtrace)$, where $TG_{\nonlogtrace}$ contains only the nodes generating $\nonlogtrace$, thus restricting the associated matrices $L$ (and $R$) into $L_\nonlogtrace$ (and $R_\nonlogtrace$). $\omega_\nonlogtrace$ is then exploited to preserving probabilities from initial (and to final) $\tau$ transitions. Such $\omega$ can be computed as $\omega := 1-\prod_{\Ind{0}\cdots\Ind{n}\in\seqs{\sigma'}{ \closed{G}}}\Big(1-(\textit{ifte}(\mathbf{1}_{L_{\sigma'}(\Ind{0})=\tau},[R_{\sigma'}]_{\Ind{0}\Ind{1}})\textit{ifte}(\mathbf{1}_{L_{\sigma'}(\Ind{n})=\tau},[R_{\sigma'}]_{\Ind{n-1}\Ind{n}})\Big)$
where $\textit{ifte}(x,y):=x(y-1)+1$ returns $y$ if $x=1$ and $1$ otherwise. %We denote the set of all the $\closed{G}_\nonlogtrace$ as $\TBf{\pmin}{4}(\closed{G})$.
%The graph weight $\omega$ derives from the outgoing edges of the initial node and the ingoing edges of the accepting node;
%such nodes are labeled as $\tau$. %Since the embedding strategy from \cite{LodhiSSCW02} considers only visible
%transitions, and the trace extraction process discards the $\tau$ information, we use $\omega$ to preserve such information.
%We call the pair consisting of a transition graph and a graph weight a \emph{weighted transition graph}.
%\begin{example}\label{ex:neue}
%\end{aligned}$$
Table \ref{tab:proj} shows the projected transition graphs associated to  traces from Example \ref{ex:rankingTaus}, where  all the $\tau$-labeled nodes are removed as required.
\begin{table}[!t]
	\caption{Projections over $\net$-traces of length $4$.}\label{tab:proj}
	\centering
	\resizebox{.3\textwidth}{!}{\begin{tabular}{>{\centering\arraybackslash} m{1cm}| >{\centering\arraybackslash} m{4cm} >{\centering\arraybackslash} m{1cm} >{\centering\arraybackslash} m{1cm} }
			\toprule
			$\nonlogtrace$&$TG_\nonlogtrace$&$l$&$\omega_\nonlogtrace$\\
			\midrule
			$\const{a}$ & \includegraphics{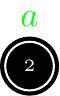} & $1$ & $\color{violet}\pa\pf$\\
			$\const{cb}$ & \includegraphics{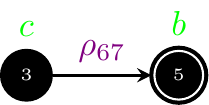} & $2$ & $\color{violet}\pb$\\
			$\const{aaa}$ & \includegraphics{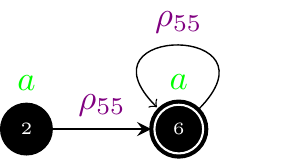} & $3$ & $\color{violet}\pa\pf$\\
			$\const{caa}$ & \includegraphics{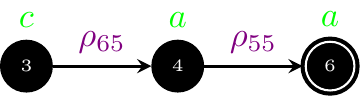} & $3$ & $\color{violet}\pb\pf$\\
			$\const{aa}$ & \includegraphics{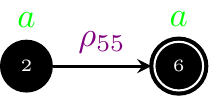} & $2$ & $\color{violet}\pa\pf$\\
			$\const{ca}$ & \includegraphics{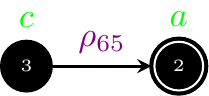} & $2$ & $\color{violet}\pb\pf$\\
			\begin{tabular}{l}aaaa\end{tabular} & \includegraphics{images/trace_a_loop} & $4$ & $\color{violet}\pa\pf$\\
			$\const{caaa}$ & \includegraphics{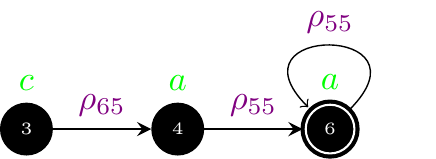} & $4$ & $\color{violet}\pb\pf$\\
			\bottomrule
	\end{tabular}}
	\vspace{-0.2cm}
\end{table}
Our proposed embedding $\gorgembed$ \cite{Bergami21} can be now exploited for each  $G_\nonlogtrace$ as follows:
%Our proposed embedding $\gorgembed$ is computed for each $wTG_\nonlogtrace$. The goal is to use
%$k_{\gorgembed}$ for ranking all the traces generated by \unravelling\ via such graphs. We extend $\trembed$ %from
%%\cite{LodhiSSCW02} including the associated probabilities, and
%by making the ranking induced by $k_{\gorgembed}$ the inverse of
%that induced by the
%sum of the following distances: the transition correlations $\epsilon$ and the transition label frequency $\nu$.
%We need that the desired properties of $\gorgembed$ are independent of the characterization of $\epsilon$ over the
%$2$-grams in $\tasks^2$ and $\nu$ over the labels in $\tasks$. %which  provide different embedding strategies. Therefore, our
%%$\gorgembed$ embedding is defined for any weighted TG.

%\begin{table}[!t]
%	\centering
%	\caption{Embedding representation for the TG $P$ in \figurename~\ref{fig:closed} and the trace $\logtrace=\textup{caba}$ after representing it as in \figurename~\ref{fig:sigmastar}. Please note that we restrict $\trace_\tau^2$ to the one from $P$.}\label{tab:emb1}
%		\begin{tabular}{l|l|l|l|l|l|l|}
%	\toprule
%	& a    & b                                                   & c    & aa   & ca   & cb   \\
%	\midrule
%	$\gorgembed(P)$ & $9.94\cdot10^{-25}$ & $1.18\cdot 10^{-26}$ & $1.04\cdot10^{-25}$ & $4.45\cdot 10^{-25}$ & $6.22\cdot10^{-25}$ & $8.29\cdot10^{-26}$\\
%	$\gorgembed(P_{\logtrace})$ & $8.16\cdot10^{-17}$ & $4.08\cdot 10^{-17}$ & $4.08\cdot10^{-17}$ & $4.37\cdot 10^{-17}$ & $1.03\cdot10^{-16}$ & $4.37\cdot10^{-17}$\\
%	\bottomrule
%\end{tabular}
%\end{table}
\begin{definition}[TG-Embedding]\label{def:ppne}
	%Given a finite set of non-empty labels $\trace_\tau =\trace\backslash\{\tau\}$, $\trace_\tau^2$ denotes all the possible pair of labels associated to paths ${\color{green}\alpha}\rightsquigarrow{\color{green}\beta}$ and $\trace_\tau$ denotes the set of all the possible non-$\tau$ node labels. Therefore, it is always possible to enumerate $\trace_\tau^2\cup\trace_\tau$ via an enumeration by a bijection $\iota\colon \trace_\tau^2\cup\trace_\tau\to  N$, where $N\subset \mathbb{N}_{\neq 0}$ and $\max N=|N|$.
	Given a weighted TG $G_\nonlogtrace=(TG_{\nonlogtrace},\omega_\nonlogtrace)$ and a tuning parameter $t_f\in[0,1]$, the \emph{TG-Embedding} is
	$$\gorgembed_{i}(G_\nonlogtrace):=\begin{cases}
	\omega_\nonlogtrace \frac{\epsilon_\const{ab}(TG_{\nonlogtrace})}{\|\epsilon\|_2}\;t_f^{|R>0|}\, & {i}=\const{ab}\in\tasks^2\\
	\frac{\nu_\const{a}(TG_{\nonlogtrace})}{\|\nu\|_2}\;\;\;\,t_f^{|R>0|}\, & {i}=\const{a}\in\tasks\\
	\end{cases}$$
	where $\nu$ and $\epsilon$ represent the embeddings associated to $L$ and $L,R$: $\epsilon$  returns
	$\epsilon_\const{ab}=0$ for the $2$-grams $\const{ab}$ not represented by the TG, and either $\nu$ always returns the empty
	vector or $\nu_\const{a}(G)=0$ iff the labels $\const{a}\in\tasks$ that are associated to no vertex in $V$.
\end{definition}
Here, ${\max\arg}_{\nonlogtrace\in \WCal{\pmin}{n}, G_\nonlogtrace\in\TBf{p}{n}(P)} k_{\gorgembed}(G_\logtrace, G_{\nonlogtrace})$ returns the best approximated trace alignment for a log trace represented as $G_{\logtrace}$. %\xout{Similarly, we can provide the TG $P\in\mathbf{P}$ providing the best approximated alignment for $P_{\logtrace}$ as $\underset{P}{\max\arg}\underset{ P_\trace\in\mathbf{P}_p^n(P)}{\max} k_{\gorgembed}(P_\trace, P_{\logtrace})$.}¯
\begin{table}[!t]
	\caption{Different sub-embeddings ($\epsilon^1$, $\epsilon^2$, $\nu^1$, and $\nu^2$) for $\gorgembed$.}\label{tab:embedstrat}
	\centering
	\resizebox{.49\textwidth}{!}
	{\begin{tabular}{c|c|c}
			\toprule
			& $x=1$ & $x=2$ \\
			\midrule
			$\epsilon^x_\const{ab}(G):=$ & $\label{eq:epsilon}
			\sum_{i=1}^l{\lambda^i}\frac{[LR^iL^t]_\const{ab}}{\sum_{\const{a'b'}\in\tasks^2}R^i_\const{a'b'}}$ & $
			\sum_{i=1}^l\lambda^i[\Lambda^i]_\const{ab}$\\
			$\nu^x_\const{a}(G):=$ & $\frac{1}{c}\sum_{\nonlogtrace'\in \ptraces{G}{0}}\frac{|\Set{\nonlogtrace_i'\in\nonlogtrace'|\const{a}\in\tasks\wedge \nonlogtrace'_i=\const{a}}|}{|\nonlogtrace'|}$ & $0$ \\
			\bottomrule
	\end{tabular}}
\end{table}
We choose two possible interchangeable definitions for $\nu$ and $\epsilon$ shown in Table \ref{tab:embedstrat}, where $l$
is the path length, and $c$ for $\nu^1$ is a normalization factor such that $\sum_{\const{a}\in\tasks}\nu^1_\const{a}(P)=1$;
$\nu^2$ completely ignores the label frequency contribution; $\epsilon^2$ is the embedding $\trembed$ from \S\ref{subsec:katk};
and $\epsilon^1$ and $\epsilon^2$ only differ from the normalization. $t_f\in [0,1]$ and $\lambda\in (0,1]$ are tuning
parameters that can be inferred from the available data. %\cite{DriessensRG06}.
The latter describes the decay factor, while $t_f$
represents the relevance of our embedding representation as the number of edges within $\closed{G}_\nonlogtrace$ increases.
We choose $t_f=0.0001$ and $\lambda=0.07$.
This representation is independent of the representation of the trace to be aligned and does not have to be
recomputed for each alignment.

%When the transition matrix is ergodic \cite{StocasticCC},  the transition matrix embedding converges to $\epsilon(R)_{\color{green}\alpha\beta}=[(\mathbf{I}-\lambda\Lambda)^{-1}]_{\color{green}\alpha\beta}$ \cite{GartnerFW03} for $n\to+\infty$.

\begin{table*}[!t]
	\centering
	\caption{Embeddings for $\net$-traces of maximum length $4$ and $\logtrace=\const{caba}$.}\label{tab:emb2}\label{tab:embsitar}
	\resizebox{\textwidth}{!}{\begin{tabular}{lllllllllllll}
			\toprule
			& $\const{a}$    & $\const{b}$                                                   & $\const{c}$    & $\const{aa}$  & $\const{ab}$   & $\const{ac}$   & $\const{ba}$  & $\const{bb}$   & $\const{bc}$   & $\const{ca}$   & $\const{cb}$  & $\const{cc}$  \\
			\midrule		
			$\gorgembed(TG_\const{aaaa})$ & $1.00\cdot 10^{-24}$ & $0$   & $0$   & $6.44\cdot 10^{-26}$ & $0$& $0$& $0$& $0$& $0$    &$0$& $0$& $0$ \\
			$\gorgembed(\overline{G}_\const{aaa})$  & $1.00\cdot 10^{-24}$ & $0$   & $0$   & $1.29\cdot 10^{-25}$ & $0$& $0$& $0$& $0$& $0$    &$0$& $0$& $0$ \\
			$\gorgembed(\overline{G}_\const{aa})$   & $1.00\cdot 10^{-24}$ & $0$   & $0$   & $2.57\cdot 10^{-25}$ & $0$& $0$& $0$& $0$& $0$    &$0$& $0$& $0$ \\
			$\gorgembed(\overline{G}_\const{a})$    & $1.00\cdot 10^{-4}$  & $0$   & $0$   & $0$   & $0$& $0$& $0$& $0$& $0$    &$0$& $0$& $0$ \\
			$\gorgembed(\overline{G}_\const{caa})$  & $7.07\cdot 10^{-25}$ & $0$   & $7.07\cdot 10^{-25}$ & $1.46\cdot 10^{-25}$ & $0$& $0$& $0$& $0$& $0$    &$2.05\cdot 10^{-25}$& $0$& $0$ \\
			$\gorgembed(\overline{G}_\const{ca})$   & $7.07\cdot 10^{-25}$ & $0$   & $7.07\cdot 10^{-25}$ & $0$   & $0$& $0$& $0$& $0$& $0$    &$1.00\cdot 10^{-8}$& $0$& $0$ \\
			$\gorgembed(\overline{G}_\const{cb})$   & $0$   & $7.07\cdot 10^{-25}$ & $7.07\cdot 10^{-25}$ & $0$   & $0$& $0$& $0$& $0$& $0$    &$0$ & $4.29\cdot 10^{-9}$& $0$ \\
			$\gorgembed(\overline{G}_\const{caaa})$ & $7.07\cdot 10^{-25}$ & $0$   & $7.07\cdot 10^{-25}$ & $1.03\cdot 10^{-25}$ & $0$& $0$& $0$& $0$& $0$    &$7.20\cdot 10^{-26}$ & $0$& $0$ \\
			\bottomrule
			$\gorgembed(\overline{G}_\const{caba})$ & $8.16\cdot10^{-17}$ & $4.08\cdot 10^{-17}$ & $4.08\cdot10^{-17}$ & $4.37\cdot 10^{-17}$ &  $0$& $0$& $0$ & $0$ & $0$    & $1.03\cdot10^{-16}$ & $4.37\cdot10^{-17}$& $0$ \\
			\bottomrule
	\end{tabular}}
	
\end{table*}
\begin{example}
	Table \ref{tab:emb2} shows the embeddings $\gorgembed(\overline{G}_\nonlogtrace)$ generated from Table \ref{tab:proj}, where the $l=|\nonlogtrace|$ for each \unravelled trace.
	After representing trace $\logtrace=\const{caba}$ as a sequence graph, we find its
	embedding $\gorgembed(\overline{G}_{\logtrace})$ with strategies $\epsilon^1$ and $\nu^1$
	{as} in Table \ref{tab:embsitar}: $\const{a}$ is the most frequent label and $\const{b}$ and $\const{c}$ are equiprobable.
	The $2$-gram $\const{ca}$ appears twice in the trace set and is more frequent than other $2$-grams.
\end{example}
These sub-embeddings satisfy the conditions required by the G-Embedding.
\begin{lemma}
	\label{lem:addedForOurPropos}
	The sub-embeddings in Table~\ref{tab:embedstrat} satisfy the requirements from Definition~\ref{def:ppne}.
\end{lemma}

The kernel $k_{\gorgembed}$ associated to $\gorgembed$ is {a function of the distance  $\|\hat{\epsilon}(G)-\hat{\epsilon}(G')\|_2^2$ and $\|\hat{\nu}(G)-\hat{\nu}(G')\|_2^2$ for traces $\logtrace$ and ${\nonlogtrace}$.}

\begin{proposition}
	\label{lem:rewritinglemma}
	Given {two weighted TGs} $(G,\omega)$ and $(G',\omega')$ with $G=(s,t,L,R)$ and $G'=(s',t',L',R')$, the definition of
	$k_\gorgembed$ is expanded to
	$\begin{aligned}
	%k_{\gorgembed}(G,G')=&
	 \omega\omega't_f^{|R>0|+|R'>0|}\left(1-\frac{\norm{\hat{\epsilon}(G)-\hat{\epsilon}(G')}{2}^2}{2}\right)+\\t_f^{|R>0|+|R'>0|}\left(1-\frac{\norm{\hat{\nu}(G)-\hat{\nu}(G')}{2}^2}{2}\right)
	\end{aligned}$
\end{proposition}

When $\hat{\epsilon}(G)$ and $\hat{\epsilon}(G')$ are affected by truncation errors
(i.e., $\norm{\hat{\epsilon}(G)-\hat{\epsilon}(G')}{2}^2\to 0$), the $\nu$ strategy intervenes as a backup ranking. The first
term of the sum does not affect the ranking, as it reduces to a constant factor.

%\xout{Given that we can now follow Definition \ref{def:ppne} for representing a trace $\trace$ as a proper embedding after transforming it as a TG $P_{\logtrace}$ (\S\ref{subsec:katk}), we can find the TG $P$ providing the best approximate match with  a trace $\trace$ as follows:}
%\[\Rcancel{\underset{{P}}{\max\arg}\;k_{\gorgembed}(P,T)}\]
%\xout{Still, this TG matching strategy does not allow to find the trace maximizing such score.} %To assess such problem, the next section is going to determine both an exact (\S\ref{subsec:exbkptap}) and an approximated strategy (\S\ref{subsec:akptap}) for probabilistically matching one single trace from the TG.
%
%\xout{Given the characterization of a TG as in \S\ref{subsec:ppn} and the embedding strategy proposed in Definition \ref{def:ppne}, We can \ADD{now} generate an embedding for each possible weighted trace $\braket{\trace,\probskip{\trace}}\in\mathcal{W}_p^n(P)$ for a given TG $P$ as described in the following definition:}

\begin{table}[!t]
	\vspace{+0.9mm}
	\caption{Comparison between the optimal ranking $\goldenrank$ and the kernel $k_{\gorgembed}$ with embedding strategies $\epsilon^1$ and $\nu^1$: arrows $\boldsymbol{\downarrow}$ remark the column of choice under which we sort the rows.}\label{tab:rank3}
	\centering
	%	\begin{tabular}{l|c|ll}
	%		\toprule
	%		$\trace$ & $k_{\gorgembed}(\trace,\logtrace)$ & \textit{kernel ranking} & expected ranking\\
	%		\midrule
	%		a & $8.16\cdot 10^{-21}$ & \textbf{1} & \textbf{\color{blue}1}\\
	%		ca & $1.89\cdot 10^{-24}$ & \textbf{2} & \textbf{\color{blue}4}\\
	%		cb & $7.64\cdot 10^{-25}$ & \textbf{3} & \textbf{\color{blue}5}\\
	%		caa & $1.14\cdot 10^{-40}$ & \textbf{4} & \textbf{\color{blue}7}\\
	%		caaa & $9.84\cdot 10^{-41}$ & \textbf{5} & \textbf{\color{blue}8}\\
	%		aa & $9.28\cdot 10^{-41}$ & \textbf{6} & \textbf{\color{red}2}\\
	%		aaa & $8.72\cdot 10^{-41}$ & \textbf{7} & \textbf{\color{red}3}\\
	%		aaaa & $8.44\cdot 10^{-41}$ & \textbf{8} & \textbf{\color{red}6}\\
	%		
	%		\bottomrule
	%	\end{tabular}
	
	\resizebox{\columnwidth}{!}{\begin{tabular}{l|ll|cc}
			\toprule
			
			{$\nonlogtrace$} &
			%\multirow{2}{*}{$d(\trace,\logtrace)$} &
			%\multicolumn{2}{c|}{$\mu_{\logtrace}$} &
			$( \probskip{\nonlogtrace}$ &  $,\,\boldsymbol{\downarrow} s_d(\logtrace,\nonlogtrace)) $ &
			{$=\goldenrank(\logtrace,\nonlogtrace)$} &
			{$k_{\gorgembed}(\closed{G}_\logtrace,\closed{G}_{\nonlogtrace})$} \\

			\midrule
			$\const{caa}$  & $0.035$ & $\;\; 0.8333$ & $0.0292$ & $1.14\cdot 10^{-40}$\\
			$\const{caaa}$  &  $0.0175$ & $\;\; 0.8333$ & $0.0145$ & $9.84\cdot 10^{-41}$\\
			$\const{a}$  & $0.4$ & $\;\; 0.6250$  & $0.2500$ & $8.16\cdot 10^{-21}$ \\
			$\const{aaaa}$  & $0.05$ & $\;\; 0.6250$ & $0.0357$ & $8.44\cdot 10^{-41}$\\
			$\const{aa}$  & $0.2$ & $\;\; 0.7142$ & $0.1428$ & $9.28\cdot 10^{-41}$ \\
			$\const{aaa}$  & $0.1$ & $\;\; 0.7142$ & $0.0714$ & $8.72\cdot 10^{-41}$\\
			$\const{ca}$  &  $0.07$ & $\;\; 0.7142$ & $0.0500$ & $1.89\cdot 10^{-24}$\\
			$\const{cb}$  &  $0.06$ & $\;\; 0.7142$ & $0.0428$ & $7.64\cdot 10^{-25}$\\
			\bottomrule
		\end{tabular}\ \begin{tabular}{l|c}
			\toprule
			
			{$\nonlogtrace$} &
			{$\boldsymbol{\downarrow}\goldenrank(\logtrace,\nonlogtrace)$} \\

			\midrule
			$\const{a}$  &  $0.2500$ \\
			$\const{aa}$  &  $0.1428$  \\
			$\const{aaa}$  & $0.0714$ \\
			$\const{ca}$  &   $0.0500$\\
			$\const{cb}$  & $0.0428$ \\
			$\const{aaaa}$  &  $0.0357$ \\
			$\const{caa}$  &  $0.0292$ \\
			$\const{caaa}$  &   $0.0145$ \\
			\bottomrule
		\end{tabular}\	\begin{tabular}{l|c}
			\toprule
			
			{$\nonlogtrace$} &
			{$\boldsymbol{\downarrow}k_{\gorgembed}(\closed{G}_\logtrace,\closed{G}_{\nonlogtrace})$} \\

			\midrule
			$\const{a}$  & $8.16\cdot 10^{-21}$ \\
			$\const{ca}$  &   $1.89\cdot 10^{-24}$\\
			$\const{cb}$  &   $7.64\cdot 10^{-25}$\\
			$\const{caa}$  &$1.14\cdot 10^{-40}$\\
			$\const{caaa}$  &  $9.84\cdot 10^{-41}$\\
			$\const{aa}$  &  $9.28\cdot 10^{-41}$ \\
			$\const{aaaa}$  & $8.44\cdot 10^{-41}$\\
			$\const{aaa}$  &  $8.72\cdot 10^{-41}$\\
			\bottomrule
	\end{tabular}}
\end{table}

\begin{example}%\small \label{ex:11}
	The products $k_{\gorgembed}(\logtrace,\nonlogtrace)=\braket{\gorgembed(\overline{G}_\logtrace),\;\gorgembed(\overline{G}_{\nonlogtrace})}$
	with sub-embedding $\nu^1$ and $\epsilon^1$, for each trace $\nonlogtrace$ appear in
	Table \ref{tab:rank3} along optimal ranking $\goldenrank$. $k_{\gorgembed}$ approximates the optimal ranking as it tends to rank the transition graphs $\closed{G}_\nonlogtrace$ (generated from $\overline{G}$ via projection) similarly to the traces over $\mathcal{R}$.
\end{example}
The kernel $k_{\gorgembed}(G,G')$ can also be expressed as a function of the dot product of the two sub-embeddings because $\tasks^2\neq\tasks$.

\vspace*{-3mm}
\begin{equation}\label{eq:corollLem1}
\begin{array}{l}\omega\omega't_f^{|R>0|+|R'>0|}\Braket{\hat{\epsilon}(G), \hat{\epsilon}(G')}+t_f^{|R>0|+|R'>0|}\Braket{\hat{\nu}(G), \hat{\nu}(G')}
\end{array}
\end{equation}

\noindent
\textbf{Properties.}\label{subsub:prop}
When two traces $\logtrace$ and ${\nonlogtrace}$ are equivalent (correspond to the same sequence of labels with the
same probability), the kernel computation reduces to $\omega\omega'$. When both weights are $1$, the kernel returns $1$.
We call this condition \textit{weak equality} because we cannot prove that when the kernel is equal to $\omega\omega'$ then the two traces are equivalent (there could be equal embeddings coming from non-equivalent traces). \figurename~\ref{fig:counterexample} show in fact two TGs generating a different set of traces but providing the same embedding.
Traces having neither $2$-grams nor transition labels in common have kernel $0$ and vice versa (\textit{strong dissimilarity}).
Due to weak equality and strong similarity, the embedding is weakly-ideal. {By previous lemmas, all combinations of sub-embeddings from Table~\ref{tab:embedstrat} give weakly-ideal $G$-embeddings.}
Last, we provide the aforementioned lemmas' statements:

\begin{figure}[!t]
	\vspace*{-0.5cm}
	\centering
	\includegraphics[scale=0.7]{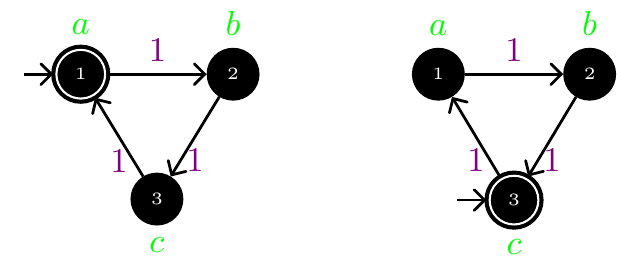}
	\caption{Two $\tau$-closed TGs, $Q$ (left) and $Q'$ (right), having a different set of traces but the same embedding.}\label{fig:counterexample}
\end{figure}
%\begin{example}
%	Using $\epsilon$ (or $\epsilon^2$) and $\nu$ (or $\nu^2$) for $\gorgembed$ may yield false positives for ``weak equality''
%	if $Q=(V,s,s,L,R)$ and $Q'=(V,s',s',L,R)$ are both cycle graphs with $s\neq s'$, $\textit{label}(s)\neq\textit{label}(s')$, and
%	$\textit{label}(s),\textit{label}(s')\neq\tau$. The graphs in Figure \ref{fig:counterexample} have the
%	same frequency for subtraces and nodes, and thus the same  $\epsilon$ and $\nu$ by construction. Using different
%	initial and accepting node with  different labels, $\ptraces{Q}{0}=\{\textup{a(bca)}^n\mid n\in\mathbb{N}\}$ and
%	$\ptraces{Q'}{0}=\{\textup{c(abc)}^n|n\in\mathbb{N}\}$. Thus, implying $\ptraces{Q}{0}\neq\ptraces{Q'}{0}$ but
%	$k_{\gorgembed}(Q,Q')=1$ for $t_f=1$.
%\end{example}

\begin{lemma}[Weak Equality]
	\label{we}
	If two weighted TGs $(G,\omega)$ and $(G',\omega')$ yield the same set of weighted traces, then
	$k_{\gorgembed}(G,G')=\omega\omega'$ for $t_f=1$.
\end{lemma}

\begin{lemma}[Strong Dissimilarity]
	\label{lem:sdiss}
	Given two weighted TGs $(G,\omega)$, $(G',\omega')$, $k_{\gorgembed}(G,G')=0$ iff $G$ and $G'$ have different set
	of vertex labels or $2$-grams with $t_f,\omega,\omega'>0$.
\end{lemma}

%% file: sections/07_experiments.tex
\begin{table}[!t]
	\caption{Distinct SWNs and associated sets of \unravelled\ traces discovered from the Sepsis Cases event log.}\label{tab:dataset}
	\centering
	\begin{adjustbox}{width=.49\textwidth}
		\begin{tabular}{crl||cl|c}
			\toprule
			\textbf{Experiment Conf.} $(\mathcal{U})$ & \textit{Model} & $+$\textit{W. Estimator} & $\pmin$& $\;\;|\WCal{\pmin}{n}(P_{\mathcal{U}})|$ \\
			\midrule
			
			\textbf{SM\_CONS\_20} &SplitMiner 2.0  \cite{AugustoCDRP19}       & +\texttt{Constant} &  $\;\;0$ & $157$  \\
			
			\textbf{SM\_FORK\_20} & SplitMiner 2.0  \cite{AugustoCDRP19}      & +Fork \cite{spdwe} &  $\;\;0$ & $32$  \\

			\textbf{SM\_PAIR\_20} & SplitMiner 2.0  \cite{AugustoCDRP19}      & +PairScale \cite{spdwe} &  $\;\;0$ & $157$ \\
			
			\textbf{STPETRI\_20} & \multicolumn{2}{c||}{Rogge-Solti \cite{RoggeSoltiAW13}} &  $10^{-5}$ & $1612$ \\
			\bottomrule
		\end{tabular}
	\end{adjustbox}
\end{table}

\section{Experimental Evaluation}\label{sec:exp}

For experimenting our proposed approach to probabilistic trace alignment, we used the Sepsis Cases event log and, from this, we generated four different datasets, where either different probabilistic weight estimations (for BPMN models) or different graph models (BPMN and our proposed SWN) were considered.\footnote{{\small \url{https://data.4tu.nl/articles/Sepsis\_Cases\_-\_Event\_Log/12707639}}} In particular, we
split the dataset into a training set, containing the ``\textit{happy traces}''  lasting at most the average trace duration in the log
($\leq 2.3\cdot 10^{7}$ ms), and a test set, containing the traces with the highest execution times. We used the training set to generate either an \uswn, using the approach presented in \cite{RoggeSoltiAW13}, or a BPMN with only exclusive gates using Split Miner 2.0 \cite{AugustoCDRP19}. Our implementation \texttt{approxProbTraceAlign}\footnote{See \url{https://github.com/jackbergus/approxProbTraceAlign} for the source code and \url{https://youtu.be/aWhS7yOa0UA} for a demo.} takes as an input both models, while internally converting the latter into a Petri net \cite{PPNFromLog}, which is later on converted into an \uswn via a firing weight estimator: we chose the \texttt{Fork} and the \texttt{PairScale} estimators from \cite{spdwe} and we denote as \texttt{Constant} a naive estimator assuming that all the transition enabled in a given marking are equiprobable. The user can select these estimators as well as the others from \cite{spdwe} from our GUI. 
%From such SWNs, we generated distinct sets of \unravelled\ traces (of different sizes): model traces can be filtered out by their minimum probability.
Finally, we loaded the testing set, which contains the log traces to be aligned against the previously loaded model (such traces can be potentially filtered in the tool). 
%by either random sampling, trace filtering by their trace length or duration, and noise can be injected to the traces with different noise degrees. In our scenario, we performed the test over the whole testing set and no noise was injected. 
Last, the GUI allows the user to tune the probabilistic trace alignment embedding by picking custom values of $t_f$ and $\lambda$.
%No estimator was used for the \uswn generated via ProM, as such engine already estimates the firing weights.
The experimental settings are summarized in \tablename~\ref{tab:dataset}. The experiments described in the following sections have the aim of evaluating the benefits of performing the approximate-ranking strategy over the optimal-ranking one.

\begin{figure*}[!t]
	\centering
	\begin{minipage}{.45\textwidth}
		\includegraphics[width=\textwidth]{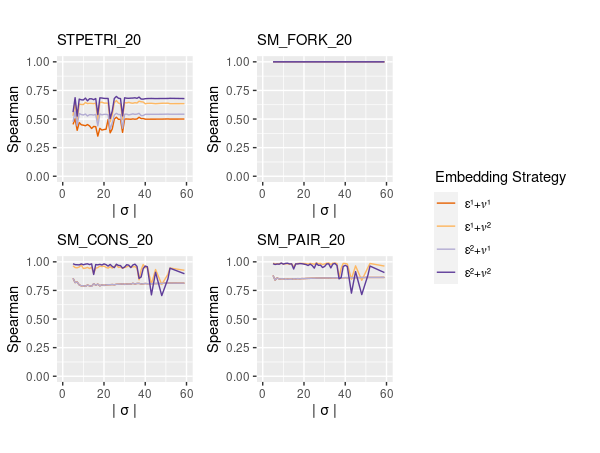}
		\caption{Approximation comparison.}\label{fig:app}
	\end{minipage}\hfill \begin{minipage}{.45\textwidth}
		
		\includegraphics[width=\textwidth]{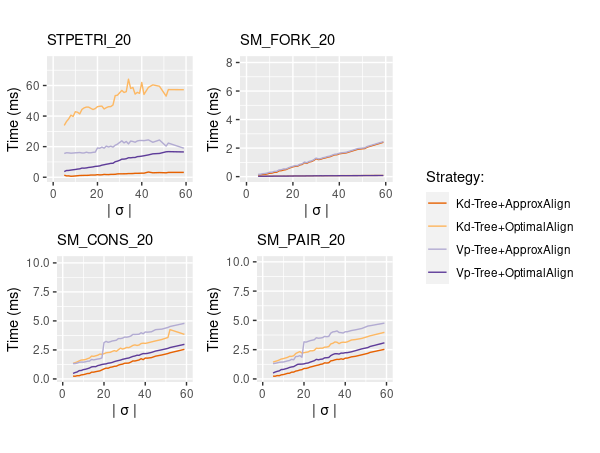}
		\caption{$k$NN alignment benchmark.}\label{fig:kronos}
	\end{minipage}
	\vspace{-0.2cm}
\end{figure*}
\subsection{Approximation}\label{subsec:apprp}
To assess how well the proposed approximate-ranking strategy approximates the optimal-ranking one, we use the Spearman correlation index \cite{BergamiBM20} to express the correlation between the ranking provided by each sub-embedding strategy for $\gorgembed$  and the optimal ranking.
\figurename~\ref{fig:app} shows the average Spearman index for traces of different lengths in the test set. We can see from the plots that the sub-embeddings considering only information about the edges (i.e., the ones where the features corresponding to the $\nu$ dimension are set to zero) have in general a higher correlation with the optimal ranking, but their correlation values are less stable w.r.t.\ the length of the trace to be aligned.
In the case of \textbf{STPETRI\_20}, the correlation is lower than for the other configurations (lower than 0.7 for all sub-embeddings). For \textbf{SM\_PAIR\_20} and \textbf{SM\_CONS\_20}, the correlation index is around 0.8 for $\epsilon^1\&\nu^2$ and $\epsilon^2\&\nu^2$, and almost 1 for $\epsilon^1\&\nu^1$ and $\epsilon^2\&\nu^1$, but less stable for these sub-embeddings especially for longer traces. In the case of \textbf{SM\_FORK\_20}, the correlation is maximum for all sub-embedding strategies.

%set of \unravelled\ traces in Table \ref{tab:dataset} and the subset of the Sepsis Cases Event Log that was not used to generate the \uswn-s. For each of this log trace $\logtrace$ we added controlled noise (transition addition, deletion, or swap) at either $20\%$ ($\tilde{\trace}^*$) or $30\%$ ($\tilde{\tilde{{\trace}}}^*$) of the log trace as for \cite{LeoniM17}. Then, we found the correlation between the ranking $R_\star$ induced by $k_{\gorgembed}(\trace,\logtrace)$ to the ranking induced by replacing $\logtrace$ with one of the two noised traces (either a ranking $R_{20}$ induced by $k_{\gorgembed}(\trace,\tilde{\trace}^*)$ or $R_{30}$ induced by $k_{\gorgembed}(\trace,\tilde{\tilde{\trace}}^*)$). The correlation $\rho$ between these two rankings ($\rho(R_\star,R_{20})$ and $\rho(R_\star,R_{30})$) is performed via Spearman Correlation Index $\rho$: such index will return near-$1$ on increasing monotonic trend, near-$(-1)$ values on decreasing monotonic trend, and near-$0$ values where the two rankings are almost uncorrelated. \figurename~\ref{fig:app} shows the outcome of such experiments for all the possible combinations of $\epsilon$ and $\nu$ sub-embeddings for $\gorgembed$ while varying the log trace length. We can observe that strategies including traces' frequencies ($\nu^1$) are more stable if compared to strategies where such information is completely ignored ($\nu^2$). Furthermore, such approximation never reaches zero values, while that occurrence might happen for $\nu^2$-based strategies.
\vspace*{-.2cm}

\subsection{Efficiency}\label{subsec:efficio}
With reference to the plots in \figurename~\ref{fig:kronos}, we evaluated the efficiency of computing the trace alignment over both optimal-ranking   and approximate-ranking   strategies over two different data structures enabling $k$NN queries, i.e., VP-Trees and KD-Trees. We conducted our experiments for $k=20$, and we used the Levenshtein distance as distance function for the optimal-ranking strategy. While the average query time (over traces of the same length) for the optimal-ranking strategy includes the \textit{indexing time} for generating all the vectors of the search space (that has to be constructed from scratch for each query) and the time for the neighborhood search, the approximate-ranking one includes the neighborhood search time and the time needed for the embedding transformation of the trace to be aligned $\logtrace$ (in this case, the indexing is performed only once before the query time); in particular, in the latter case, in addition to averaging the query time over traces of the same length, we also consider the average embedding time for all the possible embedding strategies introduced in this paper (and also used in the previous section). \figurename~\ref{fig:kronos} plots the result of such experiments: the time required to generate all the alignments needed to compute $\mathcal{R}$ truly dominates the cost of generating the embedding %$\gorgembed(\closed{G}_\logtrace)$ 
for datasets with a higher number of model traces such as \textbf{STPETRI\_20}, while the cost for embedding generation becomes non-negligible when the stochastic net generates a more restricted set of traces and, therefore, we have to compute a lower number of alignments to generate the optimal ranking (like, for example, in the case of \textbf{SM\_FORK\_20}). Finally, we can see that, in general, the computation time increases with the length of the traces to be aligned. Last, we can observe that the approximate ranking exploiting KD-Trees outperform both approximate ranking over Vp-Trees and the exact ranking with any of the aforementioned data structures. Furthermore, this configuration also provides the best trade-off between approximation and efficiency. 

%% file: sections/08_conclusions.tex
\section{Conclusions and Future Works}\label{sec:conclusion}
We tackled the probabilistic trace alignment as a $k$NN problem.
%Conceptually, this requires to handle the two possibly contrasting forces of the cost of the alignment on the one hand and the
%likelihood of the model trace with respect to which the alignment is computed. We consider the important tradeoff between both
%aspects.
The approach balances between the likelihood of the aligned trace and the cost of the alignment by providing the top-k alignments instead of a single alignment as output. The experimentation shows that the approximated top-k ranking provides a good trade-off between accuracy and efficiency especially when the reference stochastic net generates several model traces.
Future works will investigate the probabilistic alignment over fuzzy-labeled nodes and declarative process models. Also, we will try to improve the performance (in terms of efficiency and accuracy) of the proposed approach by intervening both on the embedding and the algorithmic strategies.

%\section*{Acknowledgements}
%This research has been partially supported by the project IDEE (FESR1133) funded by the Eur.\ Reg.\ Development Fund (ERDF) Investment for Growth and Jobs Programme 2014-2020. 